\documentclass[twocolumn,superscriptaddress,preprintnumbers,amsmath,amssymb,floatfix]{revtex4-1}
\pdfoutput=1
\usepackage{graphicx} % Include figure files
\usepackage{dcolumn} % Align table columns on decimal point
\usepackage{bm}% bold math
\usepackage{color}% color words
\usepackage{hyperref}% add hypertext capabilities
\usepackage{graphicx}
\usepackage{amsmath}
\usepackage{setspace}
\usepackage{appendix}

\begin{document}
\title{A Data-Driven Density Functional Model for Nuclear Systems}

\author{Zu-Xing Yang}
\affiliation{RIKEN Nishina Center, Wako, Saitama 351-0198, Japan}
\affiliation{School of Physical Science and Technology, Southwest University, Chongqing 400715, China}

\author{Xiao-Hua Fan}
\affiliation{School of Physical Science and Technology, Southwest University, Chongqing 400715, China}
\affiliation{Department of Physics, Graduate School of Science, The University of Tokyo, Tokyo 113-0033, Japan}
\affiliation{RIKEN Nishina Center, Wako, Saitama 351-0198, Japan}

\author{Zhi-Pan Li}
\affiliation{School of Physical Science and Technology, Southwest University, Chongqing 400715, China}

\author{Haozhao Liang}
\affiliation{Department of Physics, Graduate School of Science, The University of Tokyo, Tokyo 113-0033, Japan}
\affiliation{RIKEN Interdisciplinary Theoretical and Mathematical Sciences Program, Wako 351-0198, Japan}

\begin{abstract}
Through ensemble learning with multitasking and complex connection neural networks, we aggregated nuclear properties, including ground state charge radii, binding energies, and single-particle state information obtained from the Kohn-Sham auxiliary single-particle systems. 
Compared to traditional density functional theory, our model can more accurately characterize nuclear ground state information. 
Aiming at binding energy, the root mean square error is reduced to 450 keV. 
Although the complexity involving the nuclear interaction is skipped, the model has not completely devolved into a black box.
Leveraging the correlation between densities and binding energies, we calculate the neutron skin thickness of $^{208}$Pb to be 0.223 fm. 
This model will advance our understanding of nuclear properties and accelerate the integration of machine learning into modern nuclear physics.

\end{abstract}

\maketitle

\section{Introduction}

Nuclear mass is a fundamental property for extracting various nuclear structure information including nuclear pairing correlation, shell effect, deformation transition, nuclear interactions, etc \cite{Lunney2003Rev.Mod.Phys.75.10211082, Bender2003Rev.Mod.Phys.75.121180}.
In astrophysics and reaction research, nuclear mass also plays a crucial role in determining the nucleosynthesis composition on the surface of neutron stars \cite{Utama2016Phys.Rev.C93.014311} and the origin of elements in the Universe \cite{Burbidge1957Rev.Mod.Phys.29.547650}.

Considering the profound impact of mass in nuclear physics, a substantial amount of research has been devoted to enhancing the description and predictive accuracy.
Traditional theoretical models, such as the Bethe–Weizs{\"a}cker mass formula \cite{Bethe1936Rev.Mod.Phys.8.82229}, finite-range droplet model \cite{Moeller2012Phys.Rev.Lett.108.052501}, and the Weizs{\"a}cker–Skyrme model \cite{Wang2014Phys.Lett.B734.215219}, as well as the Hartree–Fock–Bogoliubov mass model \cite{Goriely2009Phys.Rev.Lett.102.152503, Goriely2009Phys.Rev.Lett.102.242501, Goriely2016Phys.Rev.C93.034337} and the relativistic mean-field model \cite{Meng2006Phys.Rev.C73.037303, Liang2008Phys.Rev.Lett.101.122502}, typically exhibit an accuracy range between 0.3 MeV and 3 MeV.
Currently, machine learning-based research is becoming gradually the main force on the path to achieving higher accuracy.
Utama et al. introduced the application of Bayesian neural networks to the residuals between theoretical and experimental data \cite{Utama2016Phys.Rev.C93.014311}, achieving remarkable success with an improvement in mass accuracy of approximately 40\%. 
The accuracy further reached an impressive 84 keV through the incorporation of nuclear pairing and shell effects \cite{Niu2018Phys.Lett.B778.4853}, coupled with meticulous design for multiple networks \cite{Niu2022Phys.Rev.C106.l021303}.
At the same time, machine learning approaches such as radial basis function \cite{Niu2013Phys.Rev.C88.024325, Niu2019Phys.Rev.C100.054311}, kernel ridge regression \cite{Wu2020Phys.Rev.C101.051301, Wu2021Phys.Lett.B819.136387}, support vector machine \cite{CLARK2006Int.J.Mod.Phys.B20.50155029}, Gaussian process \cite{Pastore2020Phys.Rev.C101.035804, Shelley2021Universe7.131, Neufcourt2019Phys.Rev.Lett.122.062502, Neufcourt2020Phys.Rev.C101.044307,Neufcourt2020Phys.Rev.C101.014319}, decision tree \cite{Carnini2020J.Phys.GNucl.Part.Phys.47.082001, Gao2021Nucl.Sci.Tech.32.}, and others were also employed to describe the nuclear masses.
From successful cases, another key insight we gather is ensemble learning, which involves integrating multiple learning models to make posterior predictions, also known as Bayesian model averaging \cite{Neufcourt2019Phys.Rev.Lett.122.062502, Neufcourt2020Phys.Rev.C101.044307, Neufcourt2020Phys.Rev.C101.014319, Hamaker2021Nat.Phys.17.14081412} or world averaging \cite{Utama2016Phys.Rev.C93.014311}.

With the description accuracy approaching the limits, the research emphasis should revert to the fundamental connections among observables for a deeper understanding of the physics behind the phenomena.
Previously, the Kohn-Sham Network (KSN) was proposed to temporarily break free from the constraints of interactions and describe the nuclear single-particle wave functions as well as the shell corrections induced by Bardeen-Cooper-Schrieffer (BCS) correlations \cite{Yang2023Phys.Lett.B840.137870}.
Under the calibration from experimental charge radii, the KSN-generated proton information received subtle adjustments, which implies that the connection between nuclear binding energies and densities needs to be re-established.

In this study, we will establish neural network mappings from the nuclear mass number density, kinetic density, and spin-orbit density to the nuclear binding energy, aiming to replenish the KSN.
We will explore network performance, generalization capability, and the impact of ensemble learning on description accuracy and outlook the future research directions.
Simultaneously, according to the correlation between densities and binding energies, the neutron skin thickness will also be further discussed.

\section{Neural network architecture \label{sec2}}

To establish the most realistic mapping relationship, we focus on two main aspects for generating inputs. 
On one hand, we derive features that empirically encompass known physical information based on proton $Z$ and neutron numbers $N$ including valence proton number $Z_v$, valence neutron number $N_v$, proton hole number $Z_h$, neutron hole number $N_h$, proton shell number $Z_s$, neutron shell number $N_s$, shell effect parameter $S$, proton number parity $Z_P$, Neutron number parity $N_P$, and parity parameter $P$.
Taking the proton as example, the relations among $Z$, $Z_s$, $Z_v$, $Z_h$, $Z_P$ satisfy
\begin{align}
    &Z = M(Z_s) + Z_v \notag\\
    &Z_h = M(Z_s+1) - Z_v \\
    &Z_P = Z \mod 2 \notag
\end{align}
with the magic number list $M$ being \{8, 20, 28, 50, 82, 126, 184\}.
The shell effect parameter $S$ and the parity parameter $P$ can be denoted as \cite{Niu2018Phys.Lett.B778.4853}
\begin{equation}
S = d_p \times d_n/(d_p + d_n) ~~~\text{and}~~~ P = [(-1)^Z + (-1)^N]/2
\end{equation}
with $d_p$ ($d_n$) representing the difference between the actual proton (neutron) numbers $Z$ ($N$) and the nearest magic number.
The $Z$, $N$, $Z_v$, $N_v$, $Z_h$, and $N_h$ share the same dimension, we denote them as $X_1$ for uniform normalization in the neural network, while $Z_s$, $N_s$, $S$, $Z_P$, $N_P$, and $P$
would be a supplement organized as $X_2$.
In addition to clearly defined shell effects and odd-even staggering, this part also plays a role in supplementing some beyond-mean-field physics that is challenging to describe within the Kohn-Sham framework, such as nucleon correlations forming on the nuclear surface, including the $\alpha$-cluster \cite{Tanaka2021Science371.260264, Typel2014Phys.Rev.C89.064321}.

On the other hand, we obtain several crucial densities from KSN single-particle wave functions $\varphi_i$ and occupancy weights $w_i$ calibrated by experimental charge radii, which are nuclear spatial density
\begin{equation}
    \rho =\sum_i \frac{d_i (\sqrt{w_i}\varphi_i)^2}{4\pi},
\end{equation}
kinetic density
\begin{equation}
    \tau =\sum_i \frac{d_i}{4\pi} \left[(\partial_r\sqrt{w_i}\varphi_i)^2+\frac{l_i(l_i+1)}{r^2}(\sqrt{w_i}\varphi_i)^2\right],
\end{equation}
and spin-orbit density
\begin{equation}
    J =\sum_i \frac{d_i}{4\pi} \left[j_i(j_i+1)-l_i(l_i+1)-\frac{3}{4}   \right]\frac{2}{r}(\sqrt{w_i}\varphi_i)^2.
\end{equation}
Here $i \in\{ {1s_{1/2}},  {1p_{3/2}},  {1p_{1/2}}, ... \}$ indicates the single-particle states, while $d_i$, $l_i$, and $j_i$ respectively represent the degeneracy, the orbital angular momentum quantum number, and the total angular momentum quantum number at a state $i$.
In density functional theory (DFT), the three aforementioned densities determine the kinetic, potential, and spin-orbit terms of nuclear interactions, thereby determining the nuclear binding energy via the Kohn-Sham equations.
In particular, when employing charge radii to calibrate densities \cite{Yang2023Phys.Lett.B840.137870}, 640 nuclei data with $Z>40$ were utilized, encompassing the majority of deformed nuclei. 
In this sense, all densities should be considered as angularly averaged. 
This implies that deformation-induced changes in binding energy should be also characterized by the features $X_1$ and $X_2$.

\begin{figure}[tb]
\includegraphics[width=8.5 cm]{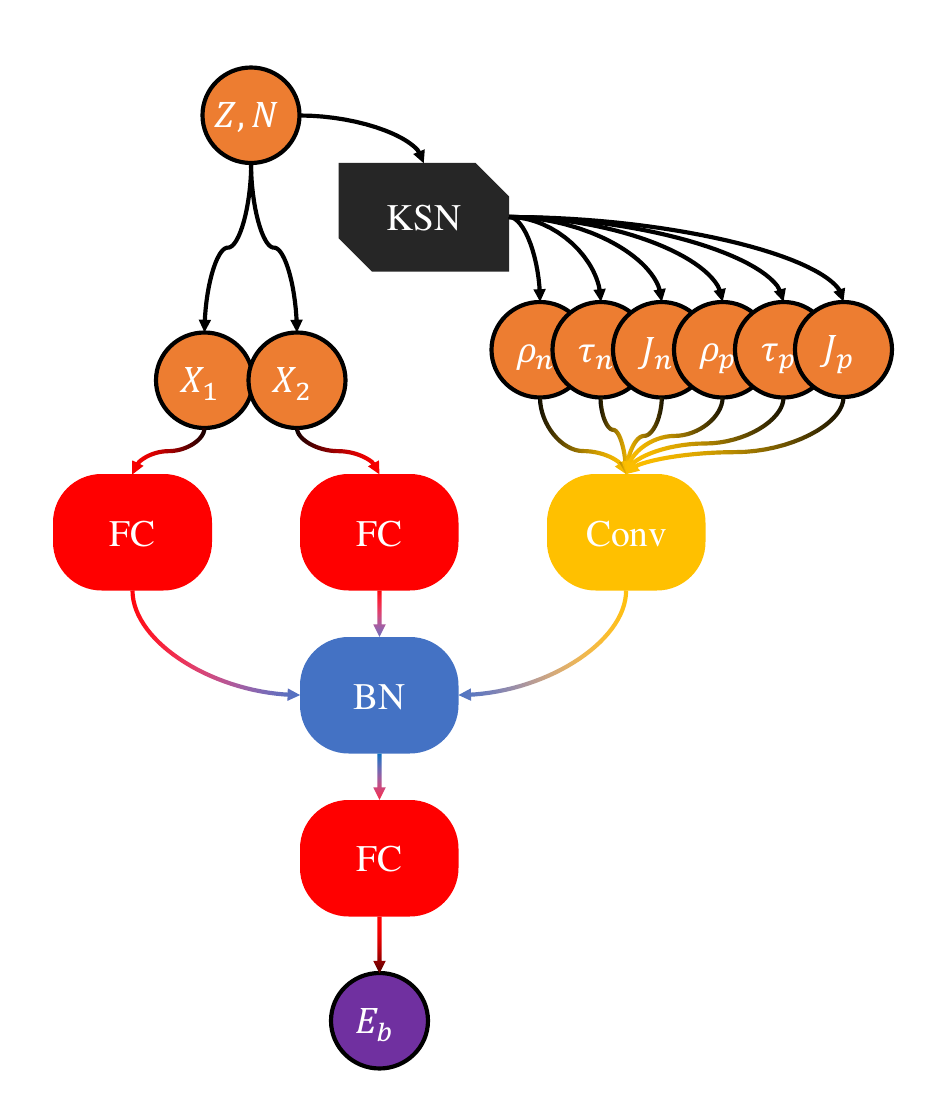}
\caption{\label{fig1}  Schematic diagram of the structure of density-to-energy network (DTEN). See the text for the abbreviation.
  }
\end{figure}

\begin{table}[tb]
\caption{\label{tab1}The hyperparameter set of DTEN structure. The ``$D$" represents the output dimension of the layer, ``$C_\text{in}$" denotes the input channels, ``$C_\text{out}$" signifies the output channels, ``$K_s$" refers to the size of the kernel which includes both convolutional and pooling dimensions, ``$S_t$" indicates the stride used during convolution or pooling operations, and ``$g(x)$" represents the non-linear activation function.}
\renewcommand{\arraystretch}{1.}
\begin{tabular}{llllllll}
\hline\hline
\multicolumn{8}{l}{$C_1$ or $C_2$}         
\\ 
\hline
L & \multicolumn{5}{l}{Type}    & $D$       & \multicolumn{1}{l}{$g(x)$}  \\  \hline
--- & \multicolumn{5}{l}{$X_1$ or $X_2$}  &     6  & --- \\
1 & \multicolumn{5}{l}{FC}  & 32      & \multicolumn{1}{l}{LReLU} \\
2 & \multicolumn{5}{l}{FC}  & 64      & \multicolumn{1}{l}{LReLU} \\
3 & \multicolumn{5}{l}{FC}  & 128      & \multicolumn{1}{l}{LReLU} \\
4 & \multicolumn{5}{l}{FC}  & 256      & \multicolumn{1}{l}{LReLU} \\
--- & \multicolumn{5}{l}{$\text{Output}_1$ or $\text{Output}_2$}   & 256      & --- \\ \hline
\multicolumn{8}{l}{$C_3$}         
\\ 
\hline
L & Type    & $D$       & $C_\text{in.}$ & $C_\text{out.}$ & $K_s$    & $S_t$    & $g(x)$      \\ \hline
--- & Densities  & (6,150)   & ---       & ---        & ---   & ---   & --- \\
1 & Conv.   & (32,150)   & 6         & 32         & 3 & 1 & LReLU \\ 
2 & pooling & (32,75) & 32         & 32          & 2 & 2 & ---       \\ 
3 & Conv.   & (64,75)   & 32         & 64         & 3 & 1 & LReLU \\ 
4 & pooling & (64,25) & 64         & 64          & 3 & 3 & ---       \\ 
5 & Conv.   & (128,25)  & 64        & 128         & 3 & 1 & LReLU \\ 
6 & pooling & (128,8) & 128         & 128          & 3 & 3 & ---       \\ 
7 & Conv.   & (256,8)  & 128        & 256         & 3 & 1 & LReLU \\ 
8 & pooling & (256,2) & 256         & 256          & 4 & 4 & ---       \\ 
9 & Conv.   & (512,1)  & 256        & 512         & 2 & 1 & LReLU \\ 
--- & $\text{Output}_3$  & 512   & ---       & ---        & ---   & ---   & --- \\
\hline
\multicolumn{8}{l}{$C_4$}   \\\hline
L & \multicolumn{5}{l}{Type}     & $D$       & \multicolumn{1}{l}{$g(x)$}  \\  \hline
--- & \multicolumn{5}{l}{$\text{Output}_1 \uplus \text{Output}_2 \uplus \text{Output}_3$}  &   1024  & --- \\
1 & \multicolumn{5}{l}{BN}  & 1024      & --- \\
2 & \multicolumn{5}{l}{FC}  & 512      & \multicolumn{1}{l}{LReLU} \\
3 & \multicolumn{5}{l}{FC}  & 256      & \multicolumn{1}{l}{LReLU} \\
4 & \multicolumn{5}{l}{FC}  & 128      & \multicolumn{1}{l}{LReLU} \\
5 & \multicolumn{5}{l}{FC}  & 32       & \multicolumn{1}{l}{LReLU} \\
6 & \multicolumn{5}{l}{FC}  & 1       & \multicolumn{1}{l}{Sigmoid} \\
--- & \multicolumn{5}{l}{Binding Energy}  & 1      & --- \\ \hline\hline
\end{tabular}
\end{table}

The mapping network from the above inputs to the nuclear binding energy is referred to as a density-to-energy network (DTEN), the structure of which is shown in Fig.~\ref{fig1}.
The $X_1$ and $X_2$ are input into two separate 4-layer fully connected (FC) neural network cells ($C_1$ and $C_2$), while the six densities  ($\rho_n$, $\tau_n$, $J_n$, $\rho_p$, $\tau_p$, $J_p$) as continuous variables are fed into a 5-layer convolutional (Conv) neural network cell with six channels ($C_3$).
Specifically, a max-pooling layer is connected after each convolutional layer to reduce the parameter and expedite convergence.
Subsequently, the outputs from these branch cells are concatenated and uniformly batch-normalized (BN) to align the feature distributions with a normal distribution. 
Afterward, passing through another FC cell ($C_4$), the features are finally mapped to the binding energy $E_b$.
An essential point that must be emphasized is that the network consists of 23 layers with intricate connections, whose complexity can lead to some neurons being trapped in the negative range and becoming deactivated under the commonly used $\text{ReLU}$ (= $\max\{0,x\}$) \cite{Villani2023.} activation function for nonlinearity, especially after multiple iterations. 
To address this issue and improve convergence, we adopt the LReLU (= $\max\{0.01x,x\}$) activation function \cite{Xu2015.} to avoid gradient vanishing.
Furthermore, to mitigate the influence of absolute data magnitudes, all input features are subject to min-max scaling, ensuring they fall within the range of 0 to 1.
Due to the constrained range of values, we employed the Sigmoid ($=1/(1+e^{-x})$) activation function in the final layer.
The meticulously designed hyperparameter set for the DTEN architecture is listed in Table~\ref{tab1}.
In the table, the initial row and the concluding row of each cell respectively signify its input and output, with the symbol ``$\uplus$" meaning concatenating two vectors, i.e.,  $(a,b,...)\uplus(c,d,...) = (a,b,...,c,d,...)$.
During the training process, a dynamically decreasing learning rate, which reduces as the loss function converges, is applied with the Adaptive Momentum Estimation (Adam) \cite{Kingma2015.}  optimizer.
The aforementioned definitions and concepts are entirely consistent with the patterns used in PyTorch \cite{PyTorchDocs}.

In the current design, non-model-dependent inputs ($X_1$ and $X_2$) and model-dependent inputs (Densities) are blended. 
The advantage is to retain accuracy as much as possible while exploring the impact of densities on binding energy.
In this manner, the search for a parameterized energy density functional is replaced, but the insights from traditional functionals are not aborted, which will produce profound significance for describing complex nuclear systems.

\section{Results and analysis \label{sec:3}}

In this study, the proton single-particle wave functions generated by KSN have undergone calibrations with the experimental charge radii of over 600 nuclei. 
However, due to the lack of neutron information from laboratories, the original SHF+BCS neutron densities with SkM* interaction \cite{Bartel1982Nucl.Phys.A386.79100} is still employed.
To ensure that the variation in density is within physically permissible limits, we need to examine the calibrated proton densities.

\begin{figure}[tb]
\includegraphics[width=9 cm]{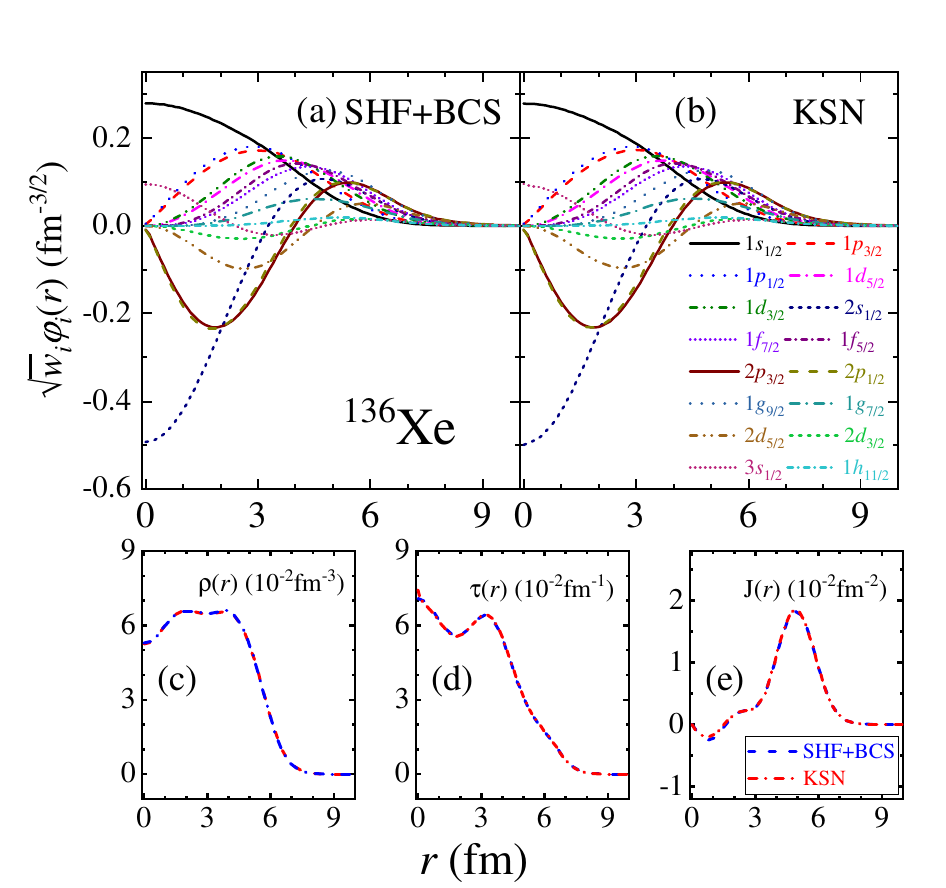}
\caption{\label{fig2} The proton $\sqrt{w_i} \varphi_i(r)$ in coordinate space for each single-particle orbital for the nucleus $^{136}$Xe with (a) SHF+BCS and (b) calibrated KSN, and the comparisons for the corresponding (c) spatial densities, (d) kinetic densities, and (e) spin-orbit densities.
  }
\end{figure}

The comparisons in Fig.~\ref{fig2} depict $\sqrt{w_i} \varphi_i(r)$ in coordinate space for each single-particle orbital, spatial densities, kinetic densities, and spin-orbit densities between SHF+BCS and calibrated KSN.
It is evident that there are only minimal changes for the various densities.
This is understandable: during the calibration process of KSN \cite{Yang2023Phys.Lett.B840.137870}, the original $\sqrt{w_i} \varphi_i(r)$ from SHF+BCS still retains a certain weight, and the calibration for charge radii often only minor changes on the typical order of a few $10^{-2}$ fm.
This implies that the loss of self-consistency in the Kohn-Sham equation is maintained at a minimal level, hence it is feasible to roughly explore changes in the kinetic, potential, and spin-orbit terms of nuclear interactions on the basis of the SkM* parameters.
However, delving into the intricate details of binding energy along such an approach directly is not practical.
As illustrated in Ref.~\cite{Yang2022.}, a neural network trained with density and binding energy data from Skyrme DFT was applied to discuss Ca isotopes, encountering a failure in describing $^{48}\text{Ca}$ with an overbinding phenomenon.
Such failure can be attributed to an indispensable beyond-mean-field effect appearing near $^{48}$Ca \cite{Perera2021Phys.Rev.C104.064313, Naito2023Phys.Rev.C107.054307}.
To capture the descriptive capacity of the beyond-mean-field effect, it is a natural approach to establish a mapping relation that targets experimental data.

\begin{figure}[tb]
\includegraphics[width=8.5 cm]{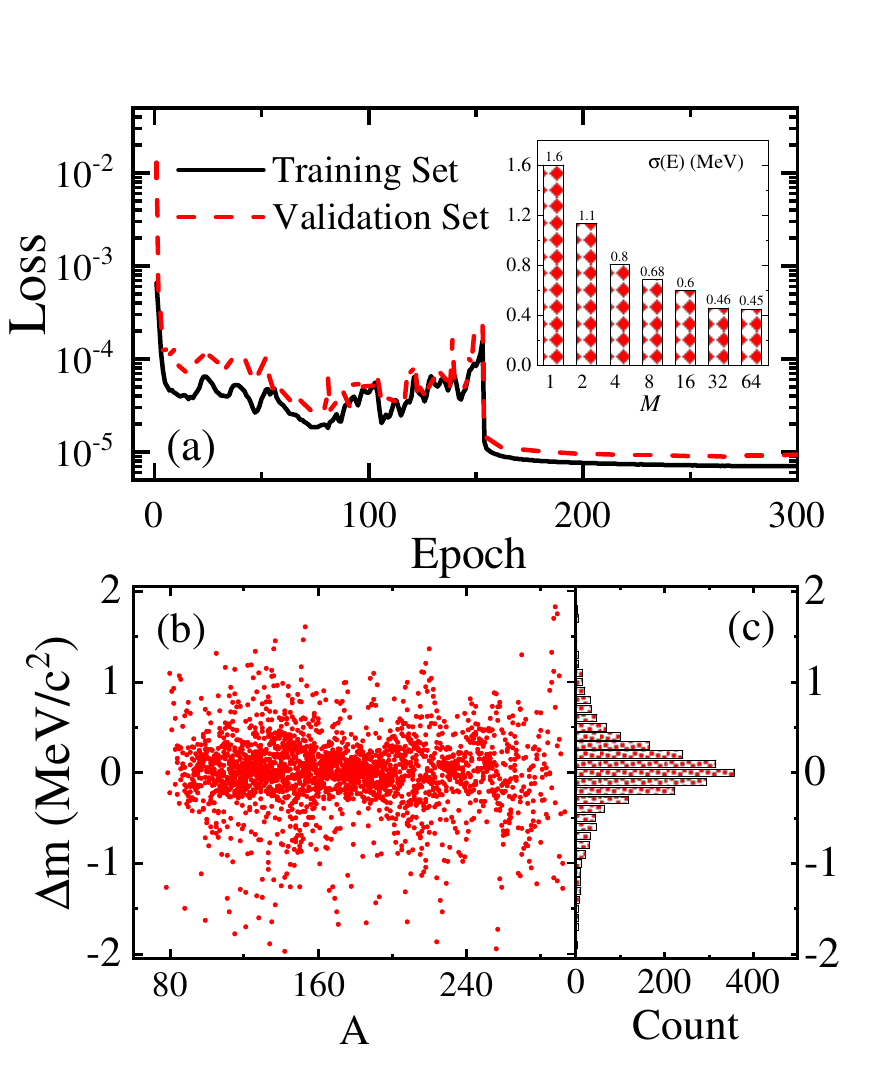}
\caption{\label{fig3} Panel (a): Loss values on the training set and validation set as a function of training epochs, where the root-mean-square binding energy error, weighted multiple models, is shown in the inner panel. Panel (b): The mass errors $\Delta m$ as a function of mass number $A$.  Panel (c): The count of mass errors in different intervals. }
\end{figure}

Turn to the training processes of DTEN, of which the changes in the loss value across training epochs are depicted in Fig.~\ref{fig3}(a).
In approximately 2400 nuclei with proton numbers greater than 40, for which binding energies have been measured with high precision, we utilized an 8:2 ratio for dividing the data into training and validation sets.
After each epoch, all nuclei in the training set have undergone training, then the decision to retain the model depends on whether there is a reduction in the observed loss value on the validation set.
Here, a weighted mean squared error is selected as the loss function,
\begin{equation}
    \text{Loss} =  \left\langle (E_{b,\text{pre}} - E_{b,\text{tar}})^2 \times A \right\rangle,
\end{equation}
where $E_{b,\text{pre}}$ and $E_{b,\text{tar}}$ represent the predicted and experimental values of the binding energy, respectively, while weight $A$ corresponds to the mass number, reflecting the higher demand for prediction accuracy in average binding energy as the mass increases.
It can be observed that the training set and validation set have essentially converged after 150 epochs, with only minor differences between them.
This indicates no overfitting occurs, emphasizing the generalization capability.
It is crucial to highlight that, in a neural network, the number of parameters typically influences training efficiency, whereas the network predictive capability is determined by the loss value and the presence of overfitting.

Although the training seems successful, upon further examination of the results, we note that the description for nuclear masses is disappointingly reflected in a root mean squared (RMS) error reaching 1.6 MeV.
To address this issue, we employ ensemble learning, combining multiple DTENs with identical structures for a final prediction.
The differences amongst these DTENs are solely caused by the random initialization parameters.
The superiority or inferiority of the network can be described by the RMS error $\sigma$, serving as a prior according to the Bayesian principle.
Therefore, the final prediction can be expressed as
\begin{equation}
    E_{b,\text{final}} = \sum_i^M w_i \times E_{b,\text{pre},i}
\end{equation}
with
\begin{equation}
    w_i = \frac{1/\sigma_i^2}{\sum_i^M 1/\sigma_i^2},
\end{equation}
where $M$ is the total number of models in the ensemble.
The inner panel of Fig.~\ref{fig3}(a) depicts the variation of the final $\sigma$ with respect to $M$.
It is clear that with the increase in the number of models, the RMS error is substantially reduced, ultimately converging to 0.455 MeV.
This precision, compared to the adopted SHF+BCS theory \cite{Reinhard1991.2850} with the RMS error being about $10\, \text{MeV}/c^2$, has been improved by order of magnitude.

As a final result, the errors of nuclear mass $\Delta m$ are demonstrated in Fig.~\ref{fig3}(b),  for which the mass number $A$ serves as the horizontal axis.
The $\Delta m$ for the majority of nuclei are within 0.5 $\text{MeV}/c^2$, while there is still a small fraction of nuclei with errors ranging from 0.5 to 2 $\text{MeV}/c^2$. 
Within this distribution, it is difficult to discern whether there is a clear correlation between the $\Delta m$ and any physical quantities.
Further projecting the errors into the counting space, as shown in (c), a standard normal distribution can be observed.
This strongly suggests that the model exhibits robustness and delivers reliable predictions.

\begin{figure}[tb]
\includegraphics[width=8.5 cm]{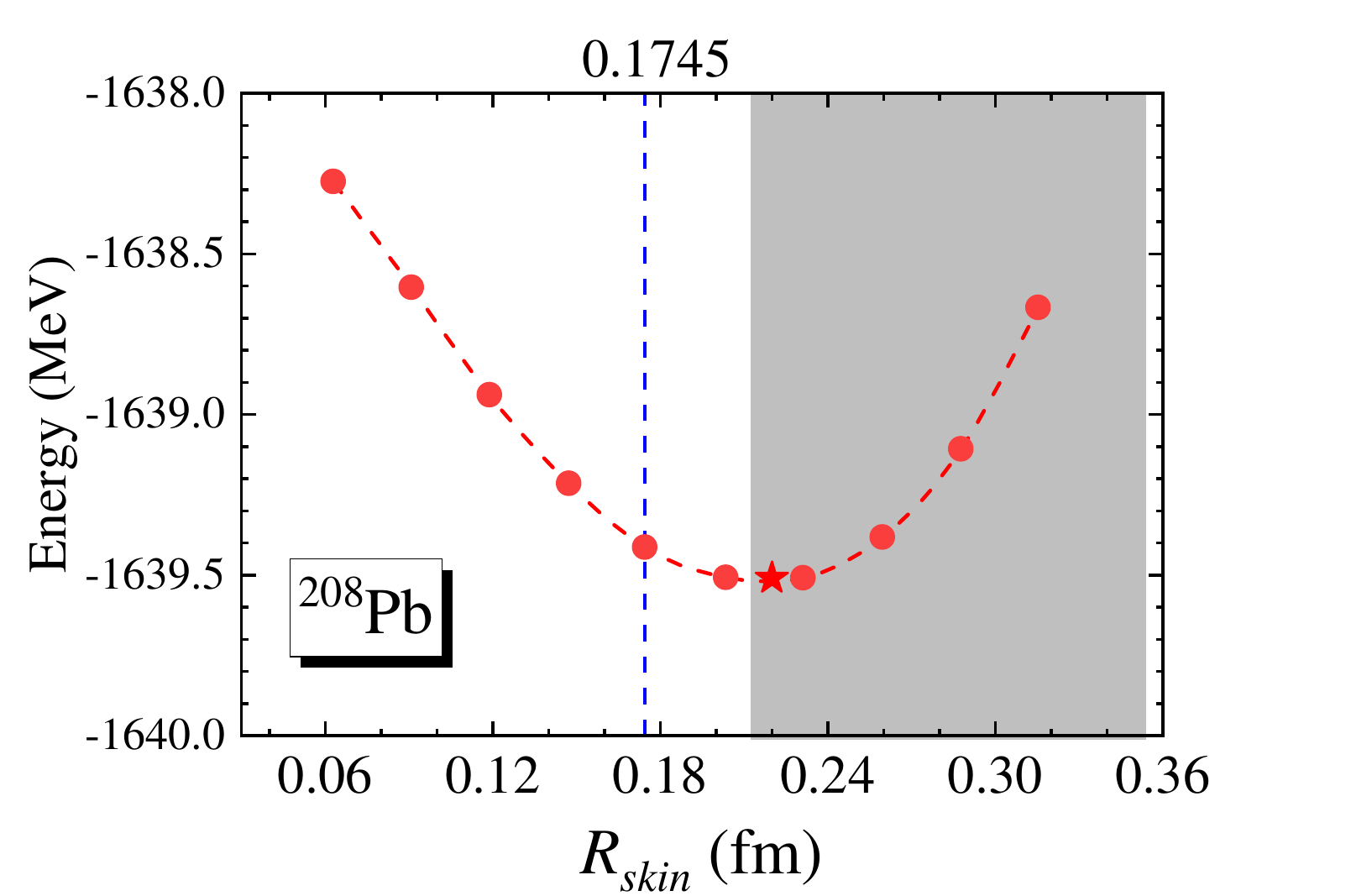}
\caption{\label{fig4} The binding energy of $^{208}$Pb as a function of neutron skin thickness. The blue dashed line represents the calculated values on the training set, while the shadow indicates the experimental value \cite{Adhikari2021Phys.Rev.Lett.126.172502}.
  }
\end{figure}

Based on the strong correlation between densities and binding energies, inferring the nuclear neutron skin thickness is an interesting and crucial problem.
The proton densities generated by KSN have been calibrated against the charge radii, and we believe they reflect the true information about protons. 
Thus, by observing the correlation between neutron density variations and the binding states of a nucleus, we can further infer the neutron skin thickness.
To this end, $^{208}$Pb, as a prominent spherical nucleus, is taken for examination.
Physically, the variation in spatial density will also lead to changes in kinetic density and spin-orbit density. 
To maintain self-consistency, based on the Kohn-Sham auxiliary single-particle system, we apply a compression operator to the neutron single-particle wave function, i.e.,
\begin{equation}
    \varphi(r) \rightarrow C_k\varphi(kr),
\end{equation}
where the parameter $k$ controls the compression ratio, while $C_k$ maintains the normalization.

With the variation of $k$, corresponding changes occur in the neutron skin and binding energy, as shown in Fig.~\ref{fig4}.
At a neutron skin thickness of 0.223 fm, the nuclear binding strength is maximized.
The obtained value is greater than the initial training value (blue dashed line) and coincidentally falls on the edge of the experimental measurement range \cite{Adhikari2021Phys.Rev.Lett.126.172502}.
Another noteworthy point is that near the minimum point, the energy changes relatively softly, less than 100 keV, suggesting that there are significant fluctuations in the neutron skin thickness.
This result is crucial for both astrophysics and nuclear reactions through the equation of state. 
We look forward to further experimental validation in the future.

The current research, in conjunction with Ref.~\cite{Yang2023Phys.Lett.B840.137870}, has essentially completed the construction of neural networks based on the Kohn-Sham scheme for enhancing DFT. 
Regarding observables, it achieves high accuracy in describing binding energies, nuclear radii, and neutron skin thickness.
In terms of physical details, it can also be employed to explore the contributions of single-particle states and shell structure, as well as the impact of various densities on binding energy.

Nevertheless, the current model still deserves further refinement and optimization. 
Primarily, when considering non-magic number nuclei, the valence-nucleon-induced deformation effects do not directly manifest in the densities. 
The current densities are assumed to be angularly averaged, which may impact the accuracy of describing binding energies and further compromise the capability to characterize multipole deformation potential surfaces.
Therefore, three-dimensionalizing the current model is imperative. 
Secondly, the current loss function may not adequately capture the differences among nuclei, especially certain beyond-mean-field effects, hindering further improving precision. 
Introducing an adversarial neural network that autonomously assesses the credibility of predictions, as opposed to a conventional RMS error, could be a more rational enhancement. 
Additionally, this may eliminate subtle non-physical zig-zag patterns technologically.

\section{summary \label{sec:4}}

With the aid of the nuclear single-particle wave functions generated by the experimental charge radius-calibrated Kohn-Sham network, we computed three essential densities in DFT, i.e., spatial density, kinetic density, and spin-orbital density. 
Through an elaborated neural network, the densities are further mapped to the experimental binding energies. 
By employing a weighted ensemble of multiple models, the RMS error in describing binding energies reached about 450 keV.
There has been a noticeable improvement compared to the initial calculations based on SHF+BCS. 
Meanwhile, the distribution of errors conforms to a standard normal distribution, reflecting the robustness of the model.

In this research, charge-radius-based calibration does not influence neutron densities. 
Therefore, it is feasible to further explore the relation between binding energy and neutron skin thickness.
Considering the self-consistency among densities, a contraction operator is applied to the neutron single-particle wave functions to establish the correlation between neutron skin and binding energy. 
By searching for the minimum point, the estimated neutron skin thickness is obtained equaling approximately 0.223 fm.

This study aggregates the charge radius data over 600 nuclei, binding energy data for more than 2400 nuclei, and single-particle state data based on DFT. 
Ultimately, it bypasses the many-body interaction potential and establishes the correlations among observables, whose descriptive performance for the nuclear ground state has surpassed that of the majority of existing density functional models.
In the future, by three-dimensionalizing the model, incorporating adversarial neural networks, as well as introducing more experimental data, the neural network for enhancing DFT will possess stronger descriptive capabilities.

\section{Acknowledgements}

This work is supported by the National Natural Science Foundation of China under Grants No.~12005175, 12375126
the Fundamental Research Funds for the Central Universities under Grant No.~SWU119076,
the JSPS Grant-in-Aid for Early-Career Scientists under Grant No.~18K13549,
the JSPS Grant-in-Aid for Scientific Research (S) under Grant No.~20H05648.
This work is also partially supported by the RIKEN Pioneering Project: Evolution of Matter in the Universe.

\bibliographystyle{apsrev4-1}
\bibliography{Ref}

\end{document}